# Contact and bulk rectification effects in ferromagnetic resonance experiments

Md. Majibul Haque Babu[1] and Maxim Tsoi[1,2,a)]

[1]Department of Physics, University of Texas at Austin, Austin, TX 78712, USA

[2]Texas Materials Institute, University of Texas at Austin, Austin, TX 78712, USA

## ABSTRACT

We present an experimental study of the spin rectification effects produced by ferromagnetic resonance in a NiFe wire. A system of four independent nonmagnetic contact probes was used to supply both rf and dc currents to the wire and to measure dc voltages at different locations in the wire. The rf current drives the ferromagnet's magnetization into resonance and produces a dc photovoltage which results from the rectification of rf current in the ferromagnet with oscillating magnetization. Our 4-probe system provided a means to detect the photovoltage and separate contributions from the ferromagnet/nonmagnet contacts and the bulk of the ferromagnet. The contact photovoltage was found to increase approximately linearly with the dc bias applied to the wire. In contrast, the bulk contribution was found to be almost independent of the dc bias. By tuning properties of individual contact probes we were able to change the magnitude of the contact photovoltage and even reverse its sign. Our results highlight the different contributions to photovoltage and the importance of contact properties/nonlinearities for rectification effects in spintronic devices.

[a)] **Author to whom correspondence should be addressed:** tsoi@physics.utexas.edu

## I. Introduction

Spin rectification [1, 2] is widely used to study magnetization dynamics in magnetic materials and spintronic devices. It refers to a dc photovoltage produced in a magnetic material or device by a rf current coupled to dynamic changes in the device resistance. The latter can be induced by various effects associated with the current, e.g., magnetoresistance [1, 2], spin-transfer torque [3], spin Hall effect [4], surface acoustic waves [5], spin-orbit interactions [6], parametric excitations [7], etc., that makes the spin rectification a versatile probe of many different phenomena. In all such experiments the rf current is injected into a magnetic material through contact electrodes (usually nonmagnetic) and propagates through the material to produce the rectification photovoltage. Thus, the photovoltage originates from both the bulk of the material as well as the contacts. In this paper we make an attempt to separate the contributions to spin rectification from the contacts and the bulk.

We study spin rectification in a $Ni_{36}Fe_{64}$ wire induced by ferromagnetic resonance (FMR), which occurs when the natural precession frequency of magnetization in the wire matches the frequency of an applied rf magnetic field generated by rf current flowing through the wire [8].

Four independent nonmagnetic (Cu) contacts were used to supply the rf current and measure rectification photovoltages at different locations in the wire. Here, the 4-point probe method gives direct access to the bulk photovoltage, while the 3-point method provides a means to measure the contact photovoltage simultaneously. The contact photovoltage was found to increase approximately linearly with the dc bias applied to the wire on top of the rf current. In contrast, the bulk contribution was found to be almost independent of the dc bias. We attribute the difference to the nonlinearity of the contacts and test this hypothesis by measuring spin rectification in different contacts. We observed that the magnitude and sign of spin rectification correlates well with nonlinearities in transport characteristics of the contacts.

## II. Methods

In our experiments we observed FMR in a ferromagnetic $Ni_{36}Fe_{64}$ wire. This is a very common ferromagnetic material, so we expect our approach to generalize to many other materials as well. A 1.2 mm long 50 μm diameter wire terminates a coaxial cable used to deliver rf and dc currents to the wire via a bias tee (see Fig. 1a). An external (dc) magnetic field $H$ is applied along the wire while the rf current produces a circumferential (rf) Oersted field perpendicular to $H$. The rf field generates a torque on the wire's magnetization and drives it into precession (FMR), thus producing a time-dependent resistance, e.g., due to anisotropic resistance of the wire. Mixing of the time-dependent resistance $R_0+\Delta R cos(\omega t+\varphi)$ with the rf current $I_{rf} cos(\omega t)$ results in a dc photovoltage

$$V_\omega = \tfrac{1}{2}I_{rf}\Delta R cos\varphi \qquad (1)$$

where $\varphi$ is the phase of the resistance changes in the wire. Because the electromagnetic skin depth in our metallic wire (~1 μm) is much smaller than its diameter, the rf current is confined to a thin layer under the wire surface, that makes our wire-like geometry equivalent to the case of a thin film in a parallel magnetic field [8]. We have experimentally confirmed [8] that FMR in our wire follows the Kittel's resonance condition for the thin-film geometry $\frac{\omega}{\gamma} = \sqrt{H(H+4\pi M_s)}$, where $\omega$ is the rf frequency, $\gamma$=280 GHz/kOe the gyromagnetic ratio, $M_s$=957 emu/cm$^3$ the saturation magnetization, and $H$ the in-plane resonance field.

The insert to Fig. 1a shows a microscope image of electrical connections to our wire. Four Cu wires with diameter of 25 μm were placed on a flat Teflon substrate under the $Ni_{36}Fe_{64}$ wire. The latter was pressed into the Cu wires by another Teflon piece controlled by a simple screw mechanism. This approach provides a means to create four (1-4) pressure-type contacts to the $Ni_{36}Fe_{64}$ wire. Contacts 1 and 4 were used to supply the rf and dc currents. All contacts were used to detect the rectified dc photovoltage produced by the rf current. The voltage $V_2$ between contacts 2 and 3 includes the bulk photovoltage, while the voltage $V_1$ ($V_3$) between contacts 1 and 2 (3 and 4) includes the contact 1 (4) photovoltage. We have also used a power sensor (Keysight U2002A) to detect the rf power reflected from the wire (not shown in Fig. 1a). This allows for a conventional detection of FMR to be used as a reference for the photovoltage detection. Both the reflected rf power and dc voltages $V_1$, $V_2$ and $V_3$ were measured as a function of $H$ at room temperature. Figure 1b shows examples of FMR spectra measured by the two methods: black trace is the photovoltage ($V_2$) and grey trace is the rf power absorption. Both methods show very similar results with FMR peak centered around a resonance field $H=\pm 1.1$ kOe. Figure 1c shows the FMR dispersion relation between the resonance field and applied rf frequency. The experiment (open circles) follows well the Kittel's FMR condition for the thin-film geometry (solid curve).

## III. Results and Discussion

Figures 2 and 3 show two examples of the photovoltages $V_1$, $V_2$ and $V_3$ measured by our 4-probe setup for two sets of contact probes 1-4. The dc resistances of contacts 1 and 4 are 6.3 and 5.8 Ω (in Fig. 2) and 21.1 and 5.8 Ω (in Fig. 3). The 4-probe resistance between contacts 2 and 3 in Fig. 2 (Fig. 3) is 0.1377 Ω (0.1376 Ω). Panels (a), (b) and (c) in Figs. 2-3 show $V_1$, $V_2$ and $V_3$, respectively, as a function of the applied magnetic field $H$ for two different dc bias currents $I_{dc}$=2 mA (grey trace) and $I_{dc}$=-2 mA (black trace). The FMR peak centered around $H=\pm 1.1$ kOe is detected in all three ($V_1$, $V_2$ and $V_3$) measurements. The bulk photovoltage ($V_2$) shows very similar results for both polarities of the applied dc bias $I_{dc}=\pm 2$ mA (see black and grey traces in Figs. 2b and 3b). In contrast, the contact photovoltages ($V_1$ and $V_3$) are inverted upon reversal of the $I_{dc}$ direction (compare black and grey traces in Figs. 2a, 2c, 3a, 3c). Also, the shape of FMR peak in $V_1(H)$ and $V_3(H)$ is strongly asymmetric; to be compared with the symmetric FMR peak in $V_2(H)$. The asymmetry can be related to a different phase of resistance variations with respect to the applied rf current [2] and will be investigated elsewhere. Here we focus on the amplitude of FMR peak measured at different dc bias currents $I_{dc}$. The dc bias dependence of the FMR peak amplitude is summarized in Figs. 2d and 3d. Here open symbols show the peak values of the bulk $V_2(I_{dc})$. Solid triangles and squares show the peak values of the contact $V_1(I_{dc})$ and $V_3(I_{dc})$, respectively. Grey lines are linear fits to the data.

The peak photovoltage in the bulk $V_2(I_{dc})$ has very little (if any) dependence on $I_{dc}$. The slopes of the linear fits to the bulk $V_2(I_{dc})$ in Figs. 2d and 3d are 0.0001 and 0.00008 V/A, respectively. These values are at the level of uncertainty of the data and much smaller than the slopes of the linear fits to the contact $V_1(I_{dc})$ and $V_3(I_{dc})$: -0.002 and -0.004 V/A (in Fig. 2d) and +0.05 and -0.005 V/A (in Fig. 3d). Note that the largest (and the only positive +0.05 V/A) slope was observed for a contact with the largest resistance (21 Ω). This contact was produced with a smaller pressure between $Ni_{36}Fe_{64}$ and Cu wires. Moreover, the current-voltage characteristic of this high-resistance contact is quite different from other contacts. Figure 4 shows the dc resistance vs bias current for the contacts in Fig. 3. The 4-probe method (bottom curve) results in a very small resistance of the bulk $Ni_{36}Fe_{64}$ wire between contacts 2 and 3 (0.1376 Ω) with essentially no dependence on the applied dc bias. The 3-probe measurements, which include the resistance of either contact 1 (top curve) or contact 4 (middle curve), show a parabolic increase or decrease of resistance with increasing bias (grey traces show parabolic fits). For contact 4 the resistance increases with the dc bias that indicates a metallic-like behavior associated, e.g., with the Joule heating. Most of the contacts in our experiments (including all contacts from Fig. 2) show this kind of behavior. However, the behavior of contact 1 in Fig. 4 is different. Here the resistance decreases with increasing bias that indicates a tunneling-like contact. The latter may be associated with an insufficient pressure between the $Ni_{36}Fe_{64}$ and Cu electrodes to break through the native oxide on wires. This produces contact with a negative (quadratic) coefficient of the parabolic resistance $\alpha$ and a positive slope of the peak FMR photovoltage $\beta$. Table 1 lists $\alpha$ and $\beta$ for all contacts (1 and 4) from Figs. 2 and 3; contacts 2 and 3 are not listed as they do not carry rf/dc currents and, thus, do not contribute to the photovoltage. We observe a good correlation between signs and magnitudes of $\alpha$ and $\beta$ coefficients.

Open symbols in Figs. 2d and 3d show that spin rectification in the bulk is virtually independent of the applied dc bias $I_{dc}$. This behavior is consistent with Eq. 1 for the spin-rectification photovoltage, where the amplitude $\Delta R$ of the time-dependent resistance is associated with the amplitude of magnetization oscillations in the wire and controlled by the rf current $I_{rf}$ (not dc current). However, spin rectification in the contacts displays a linear dependence on $I_{dc}$, which is not expected in this simple model. To explain the linear dependence, we will use the original model of rectification based on nonlinear electrical conduction [1, 9]. Here the rectification of a rf current in a nonlinear element (contact) contributes a dc photovoltage $\sim I_{rf}^2 (d^2V/dI^2)_{I_{dc}}$. The nonlinearities $(d^2V/dI^2)_{I_{dc}} \sim (d^2R/dI^2)_{I_{dc}}$ of our contacts were independently measured and are shown as $R(I_{dc})$ in Fig. 4 (for contacts from Fig. 3). Assuming a parabolic dependence of the dc contact resistance $R$ on $I_{dc}$ (see grey traces/fits in Fig. 4) these nonlinearities are $\sim \alpha I_{dc}$ and, thus, can account for the observed linear dependence of the photovoltage. One can expect a correlation between $\alpha$ and the slope $\beta$ of the photovoltage dependence on $I_{dc}$. Table 1 provides good evidence for such a correlation, including both magnitude and sign of $\alpha$ and $\beta$. The linear dependence of the photovoltage on the applied rf power ($\sim I_{rf}^2$) was independently confirmed [8] in our experiments (not shown).

## IV. Summary

We have experimentally investigated the spin rectification produced by FMR in a $Ni_{36}Fe_{64}$ wire. A system of four independent nonmagnetic (N) contacts was used to probe the rectified dc photovoltage at different locations in this ferromagnetic (F) wire. The system allowed to separate contributions to the photovoltage from the F/N ($Ni_{36}Fe_{64}$/Cu) contacts and the bulk of the ferromagnetic wire. The bulk photovoltage was found to be almost independent of an applied dc bias. In contrast, the contacts' contributions to the photovoltage were found to increase approximately linearly with the dc bias. We found a good correlation between transport properties of contacts and the magnitude and sign of spin rectification. Our results highlight the different contributions to spin-rectification photovoltage and the importance of contact properties/nonlinearities for rectification effects in spintronic devices.

This work was supported in part by the University of Texas at Austin OVPR Special Research Grant.

# References

[1] W. Egan, H. J. Juretschke, J. Appl. Phys. 34, 1477 (1963).

[2] M. Harder et al., Phys. Rep. 661, 1 (2016).

[3] J.C. Sankey et al., Phys. Rev. Lett. 96, 227601 (2006).

[4] L. Liu et al., Phys. Rev. Lett. 106, 036601 (2011).

[5] M. Weiler et al., Phys. Rev. Lett. 106, 117601 (2011).

[6] D. Fang et al., Nat. Nanotechnol. 6, 413 (2011).

[7] C. Wang, H. Seinige, M. Tsoi, J. Phys. D: Appl. Phys. 46, 285001 (2013).

[8] Q. Gao, M. Tsoi, J. Magn. Magn. Mater. 580, 170947 (2023).

[9] H. Seinige, C. Wang, M. Tsoi, Proc. SPIE **8813**, 88131K (2013).

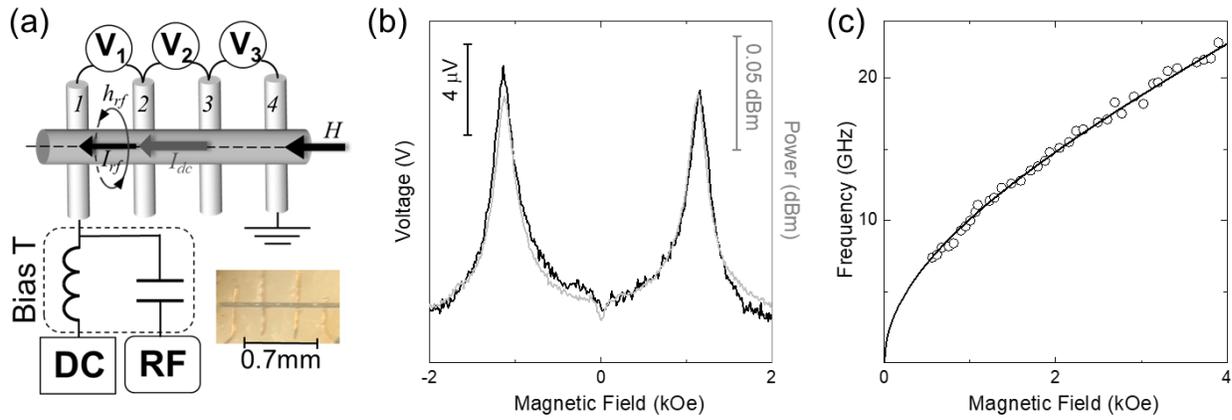

**Figure 1**. (a) Experimental setup. DC (current source and voltmeter) and RF (microwave generator and power sensor) electronics are connected to the $Ni_{36}Fe_{64}$ wire (grey; horizontal) using a bias tee. Four pressure-type contacts to the wire are made using four Cu wires (1-4; vertical). The insert shows a microscope image of the wires. (b) FMR spectra recorded by conventional absorption measurements (grey trace) and spin rectification (black trace). FMR peak is centered around ±1.1 kOe for the applied rf frequency of 11 GHz. (c) FMR dispersion (resonance field vs applied frequency). The experimental data (open circles) fitted by the Kittel's FMR condition (solid curve) for the thin-film geometry.

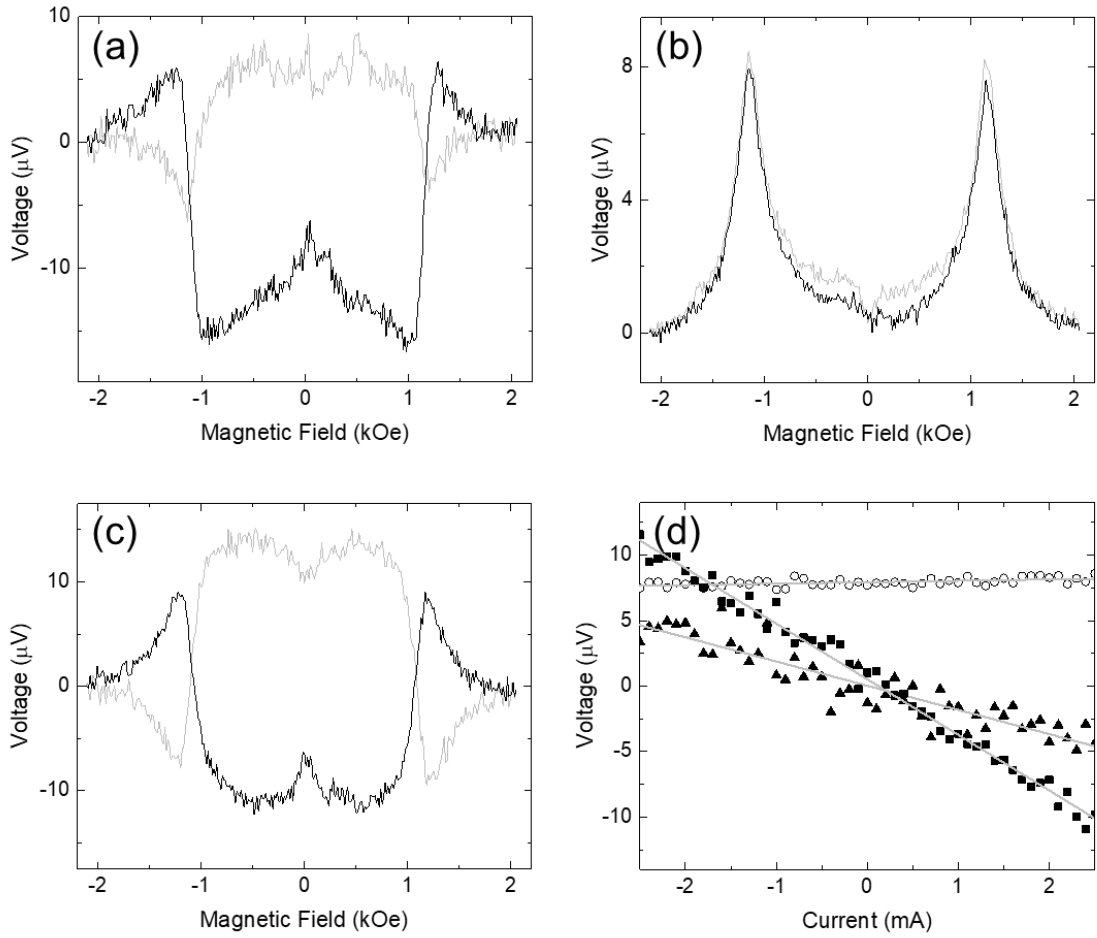

**Figure 2**. Example 1 of spin-rectification photovoltages $V_1$ (a), $V_2$ (b) and $V_3$ (c) measured as a function of the applied magnetic field $H$ for two different dc bias currents $I_{dc}$=2 mA (grey traces) and $I_{dc}$=-2 mA (black traces). The measurements were done at a constant rf frequency (11 GHz) and power (17 dBm) at the rf source. See text for details. (d) The dc bias dependence of the FMR peak amplitude. Open symbols, solid triangles and squares show the peak values of the bulk $V_2(I_{dc})$ and the contact $V_1(I_{dc})$ and $V_3(I_{dc})$, respectively. Grey lines are linear fits to the data.

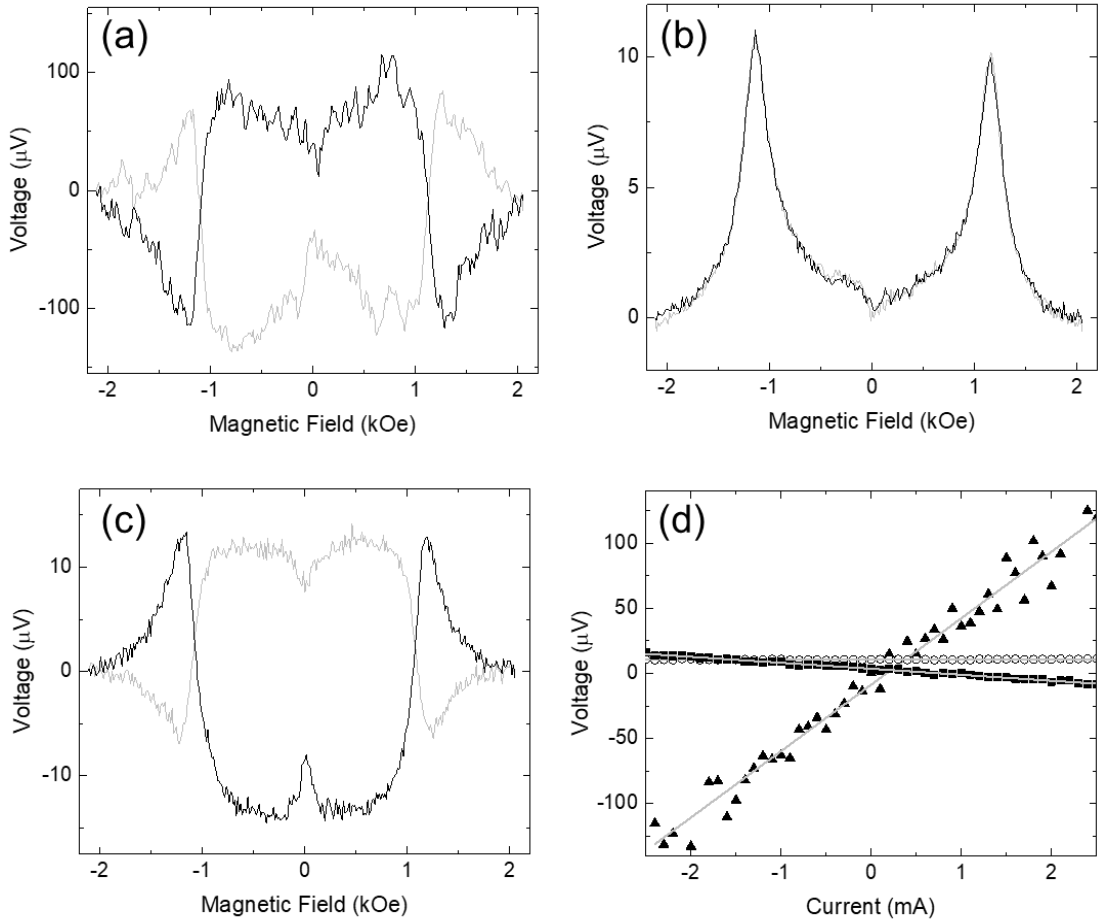

**Figure 3**. Example 2 of spin-rectification photovoltages $V_1$ (a), $V_2$ (b) and $V_3$ (c) measured as a function of the applied magnetic field $H$ for two different dc bias currents $I_{dc}$=2 mA (grey traces) and $I_{dc}$=-2 mA (black traces). The measurements were done at a constant rf frequency (11 GHz) and power (17 dBm) at the rf source. See text for details. (d) The dc bias dependence of the FMR peak amplitude. Open symbols, solid triangles and squares show the peak values of the bulk $V_2(I_{dc})$ and the contact $V_1(I_{dc})$ and $V_3(I_{dc})$, respectively. Grey lines are linear fits to the data.

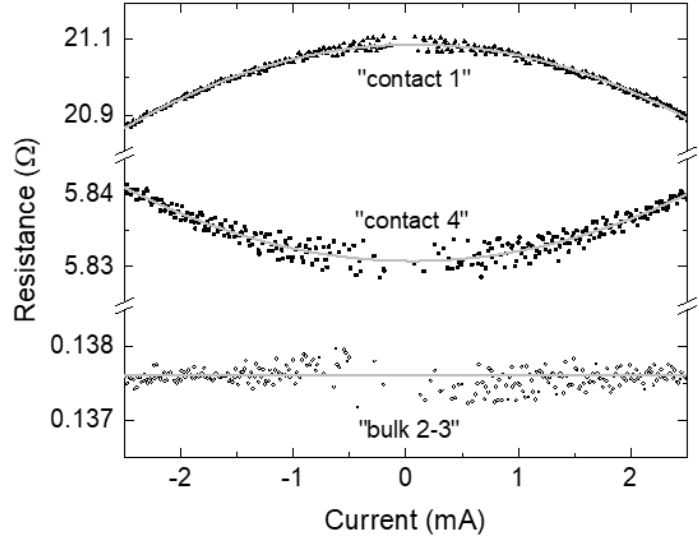

**Figure 4**. The dc resistance vs bias current characteristics of the contacts from Fig. 3. Contacts 1 (top curve) and 4 (middle curve) were measured by the 3-probe method. The bulk wire resistance between contact 2-3 (bottom curve) was measured by the 4-probe method. Symbols show the experimental data. Grey traces show parabolic fits.

| Contact # | dc resistance ($\Omega$) | $\alpha$ ($\Omega/A^2$) | $\beta$ (V/A) |
|---|---|---|---|
| #1 in Fig. 2 | 6.33 | +1140.6 | -0.00185 |
| #4 in Fig. 2 | 5.78 | +1772.4 | -0.00424 |
| #1 in Fig. 3 | 21.09 | -32549.6 | +0.05103 |
| #4 in Fig. 3 | 5.83 | +1558.7 | -0.00458 |

**Table 1**. The dc contact resistance $R=V/I_{dc}$, the quadratic coefficient $\alpha$ of the parabolic fit to $R(I_{dc})$, and the slope $\beta$ of the peak FMR photovoltage dependence on $I_{dc}$ for contacts 1 and 4 from Figs. 2 and 3.